
%
\input harvmac

\def\Titleh#1#2{\nopagenumbers\abstractfont\hsize=\hstitle\rightline{#1}%
\vskip .5in\centerline{\titlefont #2}\abstractfont\vskip .5in\pageno=0}
%
\def\UAa{Department of Physics and Astronomy}
\def\UAb{The University of Alabama}
\def\UAc{\it Box 870324, Tuscaloosa, AL 35487-0324, USA}
%
%
\def\hb{\hfil\break}

\def\tq{{\textstyle{3\over4}}}

\catcode`\@=11 

\def\lsim{\mathrel{\mathpalette\@versim<}}
\def\gsim{\mathrel{\mathpalette\@versim>}}
\def\@versim#1#2{\vcenter{\offinterlineskip
    \ialign{$\m@th#1\hfil##\hfil$\crcr#2\crcr\sim\crcr } }}
\def\boxit#1{\vbox{\hrule\hbox{\vrule\kern3pt
      \vbox{\kern3pt#1\kern3pt}\kern3pt\vrule}\hrule}}

\def\cl{\centerline}
\def\etal{{\it et al.}}

\def\t1{{\tilde 1}}

\def\tg{{\tilde g}}
\def\tp{{\tilde \gamma}}
\def\tq{{\tilde q}}
\def\tf{{\tilde f}}

\def\a3{{\alpha_3}}

\def\GeV{\,{\rm GeV}}

\def\NPB#1#2#3{Nucl. Phys. B {\bf#1} (19#2) #3}
\def\PLB#1#2#3{Phys. Lett. B {\bf#1} (19#2) #3}

\def\PRD#1#2#3{Phys. Rev. D {\bf#1} (19#2) #3}
\def\PRL#1#2#3{Phys. Rev. Lett. {\bf#1} (19#2) #3}
\def\PRT#1#2#3{Phys. Rep. {\bf#1} (19#2) #3}

\def\UAHEP#1{Alabama preprint UAHEP#1}
%
%
\nref\CCY{L. Clavelli, \UAHEP{921}, (Phys. Rev. D in press);\hb
L. Clavelli, P.W. Coulter, and Kajia Yuan, \UAHEP{924}.}
\nref\BGM{R. Barbieri, L. Girardello, and A. Masiero, \PLB{127}{83}
{429}.}
\nref\BH{R. Barbieri, L.J. Hall, \PRL{68}{92}{752}.}
\nref\HC{H. Baer, X. Tata, and J. Woodside, \PRD{44}{91}{207};\hb
UA2 Collaboration, J. Alitti \etal, \PLB{235}{90}{363};\hb
CDF Collaboration, F. Abe \etal, \PRL{62}{89}{1825},
{\bf 68} (1992) 447.}
\nref\DataG{Particle Data Group, Phys. Rev. D {\bf 45}, No. 11 (1992).}
\nref\AMP{G. Altarelli, B. Mele, and R. Petronzio, \PLB{129}{83}{456};\hb
H.D. Dahmen, D. Schiller, and D. Wahner, \NPB{227}{83}{291}.}
\nref\HEB{T. Hebbeker, Aachen preprint PITHA 91/17, (to appear in
the Proceedings of the LEP-HEP 91 Conference, Geneva, 1991).}
\nref\LEP{The LEP Collaborations, \PLB{276}{92}{247}.}
\nref\ASP{ASP Collaboration, C. Hearty \etal, \PRD{39}{89}{3207}.}
\nref\TOP{J. Steinberger, \PRT{203}{91}{345}.}
\nref\CDF{CDF Collaboration, F. Abe \etal, \PRD{45}{92}{3921}.}
%
\nfig\I{Values of $\delta$ as a function of squark and gluino mass.
Contours of constant $\delta$ are plotted with the value of $\delta$
shown to the right of the contour. It is assumed that all squark masses
are equal.}
\nfig\II{Values of $\delta$ as a function of $m_0$ and the gluino mass
for ${\cos}2\beta=0.5$. Contours of constant $\delta$ are plotted
with the value of $\delta$ shown to the right of the contour.}
\nfig\III{Values of $\delta$ as a function of $m_0$ and the gluino mass
for ${\cos}2\beta=-0.5$. Contours of constant $\delta$ are plotted
with the value of $\delta$ shown to the right of the contour.}
\nfig\IV{Values of $\delta$ as a function of $m_0$ and ${\cos}2\beta$
for zero gluino mass. That part of the graph to the right of the dashed
lines is the region allowed by the restriction that the squark masses
are above $M_Z/2$. The two enclosed areas are the regions allowed
by Eq. \XII\ and the restrictions that the slepton mass must be above
65 GeV and the sneutrino mass must be greater than 41 GeV.}
\leftline{\titlefont THE UNIVERSITY OF ALABAMA}
\Titleh{\vbox{\baselineskip12pt\hbox{UAHEP9210}}}
{\vbox{\cl{Squarks Below the $Z$}}}
\cl{L. CLAVELLI, P.W. COULTER, B. FENYI,}
\cl{C. HESTER, PETER POVINEC, KAJIA YUAN}
\bigskip
\bigskip
\cl{\UAa}
\cl{\UAb}
\cl{\UAc}
\bigskip
\bigskip
\bigskip
\bigskip
\cl{ABSTRACT}
\bigskip
We investigate the possibility that the difference between the
measurements of $\a3(M_Z)$ from the hadronic branching ratio of the
$Z^0$ and the world average of other measurements is due to the decay
of the $Z^0$ into quark, anti-squark, and gluino. Consequences for
supersymmetry breaking models are discussed.
\bigskip
\Date{July, 1992}

Recently, some of us have pointed out several features of quarkonium
decay \CCY\ suggesting the existence of gluinos in the sub-GeV mass
region. Gluinos below $1\GeV$ are natural in the $m_{1/2}=0$ model
of soft SUSY breaking, which corresponds to the low-energy manifestation
of spontaneously broken $N=1$ supergravity with minimal gauge kinetic
term \BGM. In this model, the superpartners of the massless gauge
bosons are themselves massless at tree level receiving small calculable
masses in perturbation theory, and SUSY breaking is dominantly
driven by a universal scalar squared mass $m^2_0$ representing the
difference between the average squared masses of the fermions and
their superpartners.  Assuming degenerate super-heavy particles at
the GUT scale, minimal SUSY unification with light gluinos and the
world average values of the standard model parameters require \CCY\
at the one standard deviation level that
\eqn\I{75\GeV<m_0<270\GeV .}
Non-degenerate super-heavies could loosen this prediction \BH.
Assuming negligible left-right mixing for the light quark partners,
the lower limit in Eq. \I\ insures that none of the sfermions except
perhaps the lightest stop quark is below half the $Z^0$ mass.
The $M_Z/2$ experimental lower
limit on the stop quark constrains the other parameters of the
$m_{1/2}=0$ model. This entire range of $m_0$ leads naturally in the
$m_{1/2}=0$ model to gluino masses in the sub-GeV region.
Values of $m_0$ near the lower end of this allowed range are
characterized by squark masses between $M_Z/2$ and $M_Z$ and it is
the purpose of this paper to explore the consequences of this
possibility. Contrary results from hadron colliders \HC\
suggesting $100\GeV$ level lower bounds for the masses of squarks
and gluinos are dependent on the assumption that these particles,
once produced, will lead to significant amounts of missing transverse
energy due to their ultimate decay into photinos or other weakly
interacting stable neutralinos. The hadronization Monte Carlos
leading to the quoted bounds are reliable at some level for gluino
masses in the multi-GeV or tens of GeV range but are presumably not
intended to be used for light gluinos. The following plausible
scenario assuming squark masses above half the $Z^0$ mass and a
gluino mass below one GeV would invalidate the hadron collider limits.
A squark produced in hadronic collisions will immediately decay into
the corresponding quark plus a gluino. The decay gluino or a gluino
independently produced in the hadronic collision would hadronize into
a heavy jet of hadrons one of which would ultimately be the lightest
hadron with a gluino constituent. Such a hadron would presumably be
in the 1.5 to 2.5 GeV mass region as expected for gluon bound states.
The gluino-containing hadron would decay predominantly into a
multi-pion final state plus a (predominantly soft) photino.
Such a photino would very rarely carry off a significant amount of
transverse energy. Thus if the gluinos are in the sub-GeV region the
major effect in hadronic collisions will be only a somewhat higher
apparent value of the strong coupling constant with little difference
in the jet topology. In this scenario the tightest limits on SUSY
particles will come from ${\rm e}^+{\rm e}^-$ colliders.

It is well known that the non-observation of the decay of the $Z^0$ into
squark-antisquark pairs sets a lower limit of approximately $42\GeV$
for the lightest squark mass \DataG. However, if the gluino is much
lighter than $Z^0$, a stronger bound can be put on the squark mass
due to the effect on the $Z^0$ hadronic branching ratio of the decay
\eqn\II{Z^0\rightarrow {\bar q}\tq\tg + {\rm Charge}\ {\rm Conjugate}}
where the tildes indicate supersymmetric partners of the quark $q$ and
gluon $g$.

The effect of squarks and gluinos on ${\rm e}^+{\rm e}^-$
annihilation into hadrons through a virtual photon was treated a decade
ago \AMP\ and these results can be carried over simply to $Z^0$ decay.
The hadronic to electronic branching ratio of the $Z^0$ is of the
form \HEB\
\eqn\III{{\Gamma(Z^0\rightarrow {\rm hadrons})\over
          \Gamma(Z^0\rightarrow {\rm e}^+{\rm e}^-)}
          =19.97\Bigl[1+1.05{\a3(M_Z)\over \pi}
           +0.9\Bigl({\alpha_3(M_Z)\over \pi}\Bigr)^2
    -13\Bigl({\alpha_3(M_Z)\over \pi}\Bigr)^3 + \cdots\Bigr].}
Including the effect of virtual squark-antisquark pairs and
quark-squark-gluino decay channels, the lowest order effect of the SUSY
particles is to replace $\a3$ by $\alpha_{\rm eff}$ in the linear term
of Eq. \III.
\eqn\IV{\a3(M_Z)\rightarrow \alpha_{\rm eff}(M_Z)=\a3(M_Z)(1+\delta)}
The $\delta$ parameter, which is a function of the squark and gluino
masses, gives the lowest order effect of the SUSY particles assuming
the squark mass is above half the $Z^0$ mass. Since the effects of
SUSY particle on the other terms are not readily available we make the
substitution of Eq. \IV\ in the quadratic and cubic terms of Eq. \III\
also, accepting a theoretical error of order $(\a3(M_Z)/\pi)^2\delta$.
$\delta$ can be calculated from Eqs. (3) and (8) of the first paper of
Ref. \AMP\ as a function of the squark mass and the gluino mass.
The typographical error in the last term of Eq. (8) has been corrected
in our calculations. In these equations the quark mass is neglected
by comparison to $M_Z$ and $m_\tq$.
The results for $\delta$ are presented in Fig.~1 including
all quarks except the top and setting all squark masses equal as in
Ref. \AMP. Results for more realistic calculations in which the squark
masses differ are discussed below. The effective strong coupling
constant as deduced by the experimental value of the left hand side
of Eq. \III\ is \LEP\
\eqn\V{\alpha_{\rm eff}(M_Z)=0.141\pm 0.017}
which can be compared with the world average value \HEB\
\eqn\VI{\a3(M_Z)=0.113\pm 0.003}
At the one standard deviation level, Eq. \IV\ would imply that
\eqn\VII{\delta=0.25\pm 0.15}
One could, conservatively, take Eq. \VII\ as putting an upper limit
of 0.4 to the contribution of squarks and gluinos to $\delta$.
We would not expect an appreciable fraction of this to be attributable
to the higher order process $Z^0\rightarrow q{\bar q}\tg\tg$. From
Fig.~1, assuming negligible gluino mass, this translates into an
experimental bound
\eqn\VIII{m_\tq > 50\GeV}

Since, however, in the $m_{1/2}=0$ scenario for soft SUSY breaking the
gluinos are expected to be below 1 GeV and the squarks could be between
$M_Z/2$ and $M_Z$, it is interesting to consider the consequences of
the assumption that squarks and gluinos are in fact causing the
deviation of $\delta$ from zero.  Arguing against this interpretation
is the fact that the $R$ parameter of ${\rm e}^+{\rm e}^-$ measured
at 35 GeV where there is certainly no squark contribution also yields
an anomalously large value of $\a3$ (with large errors) \HEB.
We might suggest, however, that at 35 GeV there is room for higher
twist, power law, contributions to the $R$ parameter at the level of
a few percent. This would be sufficient to bring the value at 35 GeV
into line with other measurements. At the $Z^0$ mass these should be
negligible and therefore one might expect the hadronic decay rate of
the $Z^0$ to accurately reflect the perturbative QCD final states.
Eq. \VII, at the one standard deviation level, together with Fig. 1
would then imply
\eqn\IX{50\GeV <m_\tq< 82\GeV}
As discussed above, the uncertainties in the hadronization models
built into the jet Monte Carlos are in the case of light gluinos
such that Eq. \IX\ can probably not be reliably ruled out by the
present hadronic collider data. The sfermions, except for the stop
quark, are expected in a good approximation to be unmixed partners
of the left and right handed fermions with masses
\eqna\X
$$\eqalignno{m^2_{\tf ,LL}&=m^2_0+m^2_f+M^2_Z{\cos}2\beta\Bigl[
            T_{3,f}-e_f{\sin}^2\theta_W\Bigr],&\X a\cr
m^2_{\tf ,RR}&=m^2_0+m^2_f+M^2_Z{\cos}2\beta
e_f{\sin}^2\theta_W.&\X b\cr}$$
The average squark squared mass as well as the average slepton squared
mass therefore satisfies
\eqn\XI{\langle m^2_\tf \rangle =m^2_0+\langle m^2_f \rangle .}
In the $m_{1/2}=0$ scenario the angle $\beta$, the arc-tangent of the
ratio of Higgs vevs is constrained by current data to lie in one of
two ranges \CCY\
\eqn\XII{-0.674<{\cos}2\beta <-0.385\quad {\rm or}\quad
0.406<{\cos}2\beta <0.536}
Some squark masses in the range of Eq. \IX\ would follow in this
scenario if $m_0$ were near the lower end of the range in Eq. \I.
Because of the non-linearity of $\delta$ as a function of the
squark masses, the effective mass in Eq. \IX\ for determing $\delta$
lies between the lightest and the average squark mass.

We ask the question: If a $\delta$ in the range of Eq. \VII\ is to be
produced by squarks and gluinos what must be the values of
$m_0$ and $\beta$ and what would be the consequences for slepton
and top quark masses? If these consequences can be ruled out
experimentally alternate interpretations must be sought for the $\delta$
anomaly. Of course the anomaly is only a two standard deviation effect
which could easily disappear with better statistics. In either case the
parameters of the $m_{1/2}=0$ model would be further constrained with
important implications for the gluino and squark masses.

{}From Eq. \X{}\ we can see that, if ${\cos}2\beta$ is positive,
the ``down-type" squarks ($T_3=-1/2$) would be below $m_0$
and the ``up-type" squarks would be above $m_0$, with the
reverse situation holding if ${\cos}2\beta$ were negative.
The squared coupling to the $Z^0$ is proportional to
$T^2_{3,f}+(T_{3,f}-2e_f{\sin}^2\theta_W)^2$ which is greater for
``down-type" squarks than for ``up-type". Thus, for fixed $m_0$,
a greater contribution to $\delta$ would result if
${\cos}2\beta$ were positive. However, current experimental constraints
are such that lower $m_0$ is allowed if ${\cos}2\beta$ is negative so
that the largest effect consistent with current data is obtained
if ${\cos}2\beta$ is negative and is dominated by the effect
of ``up-type" squarks. In Figs. 2 and 3 we show the $\delta$ contours
as functions of $m_0$ and the gluino mass in the cases of positive and
negative ${\cos}2\beta$ respectively. However, apart from the window at
very low gluino mass, most of the parameter space of Figs.~2 and 3 is
ruled out by collider and beam dump searches for SUSY particles.
In Fig.~4 we show the $\delta$ contours as a function of $m_0$
and ${\cos}2\beta$ assuming negligible gluino mass as suggested in
the second article of Ref. \CCY. The currently allowed regions
of parameter space in the $m_{1/2}=0$ model with $m_0<95\GeV$
lie in the trapezoid in the upper right of Fig.~4 and in the pentagonal
region in the lower right. The region to the left of the slanted lines
of postive slope is ruled out by the current lower limit of 65 GeV on
the mass of the selectron \refs{\ASP,\DataG} while the region to the
left of the
slanted line of negative slope is ruled out by the current limit on
sneutrino masses. As can be seen from Fig.~4, squarks below the $Z^0$
can account for the anomalously large apparent value of
$\a3(M_Z)$ seen in $Z^0$ decay if $m_0$ is in the range
\eqn\XIII{60\GeV<m_0<85\GeV}
with values above $\delta=0.16$ allowed only for negative values of
${\cos}2\beta$.

Such a low value of $m_0$ would also have strong consequences for the
top quark mass since minimal SUSY unification with gluinos below the
$Z^0$ leads to the effective SUSY mass \CCY\
\eqn\XIV{M_S=150\GeV\times e^{-518.5({\sin}^2\theta_W-0.2333)}
e^{1.85(\a3^{-1}(M_Z)-0.113^{-1})}.}
This equation with the world average values of $\a3$ and
${\sin}^2\theta_W$ leads to Eq. \I. However, since we are exploring
the consequences of light SUSY particles, for $\a3(M_Z)$ we use the
tightly constrained value coming from the best fit to
the quarkonia data allowing for light gluinos \CCY,
$\a3(M_Z)=0.113\pm 0.001$. Then requiring an $M_S$ consistent with
Eq. 13 leads to a tight prediction for ${\sin}^2\theta_W$.
The experimental result for ${\sin}^2\theta_W$ is strongly dependent
on the top quark mass \TOP\
\eqn\XV{{\sin}^2\theta_W(M_Z)=0.2327\pm 0.0004+0.0043\Bigl[
1-{m_t\over 125\GeV}\Bigr].}
The quoted error here comes in equal parts from the experimental error
and the uncertainty due to the Higgs mass which is assumed to lie
between 50 GeV and 1 TeV. It is clear from Eqs. \XIV\ and \XV\ that
lower values of $m_t$ are correlated with lower values of $M_S$.
\eqn\XVI{m_t=\Bigl[125+56.1\ln\Bigl({M_S\over 205}\Bigr)\pm
14\Bigr]\GeV .}
The observation of Ref. \BH\ would increase the error here unless the
GUT-scale particles are very nearly degenerate. Equating $M_S$
with the average squark mass squared leads via Eq. \XI\ to
\eqn\XVII{M^2_S=m^2_0+1608{\GeV}^2}
where, in calculating the average quark mass, we have used for $m_t$
a value near the current experimental lower limit since in order to
bring $m_0$ into the range of Eq. \XIII, one needs small values of
the top quark mass. As $m_0$ increases through the range of Eq. \XIII,
the top quark mass runs from $67\pm 14$ GeV to $81\pm 14$ GeV. These
values are very close to the best fit values of $m_t$ arising from the
the total unconstrained LEP data \LEP. However only the highest values
here are consistent with the recent CDF limit \CDF\ $m_t>91\GeV$.
Although the possibility of SUSY decays might loosen the CDF limit,
their result taken at face value would disfavor a value of
$\delta$ greater than 0.1 coming from the mechanism discussed here
which however is sufficient to bring $\alpha_{\rm eff}(M_Z)$ into
agreement with
the world average value $\a3(M_Z)$ at the one standard deviation
level. This would require a value of $m_0$ between 75 and 85 GeV in
agreement with the lower limit of Eq. \I.

The following conclusions can be drawn from our analysis. The current
experimental constraints on the $m_{1/2}=0$ model of
SUSY breaking are such that the discrepancy between the
$\a3(M_Z)$ value from the $Z^0$ width and the world average is
consistent (at the one standard deviation level) with an explanation
in terms of $Z^0$ decay into quark-antisquark-gluino.
This explanation is however not a prediction of the $m_{1/2}=0$
model since it requires that $m_0$ be near the bottom of the currently
allowed range. If this explanation is in fact chosen by nature, many of
the SUSY particles are below the $Z^0$ and will be clearly seen
at LEP II. The non-collinear lepton pairs
$Z^0\rightarrow \mu^+\mu^-\tp\tp$
coming from the decay
$Z^0\rightarrow \mu^+{\tilde \mu}^-\tp+{\rm C.C}$
should be seen even at LEP I when increased statistics are available.
In addition the top quark should then be at the very
low end of the currently allowed range and should
be seen in future Fermilab experiments.

\bigskip
\bigskip
\cl{\bf Acknowledgments}\nobreak
This work has been supported in part by the U.S. Department of Energy
under Grant No. DE-FG05-84ER40141 and by the Texas National
Laboratory Research Commission under Grant No. RCFY9155.
\listrefs
\listfigs
\bye